\documentclass[useAMS,usenatbib]{mn2e} 
\usepackage{aas_macros}
\usepackage{graphics}
\usepackage[pdftex]{graphicx}
\usepackage{epstopdf}
\usepackage{epsfig}  
\usepackage{natbib} 
\usepackage{dingbat}
\usepackage{float}
\usepackage{amsmath}
\usepackage{times}
\usepackage[varg]{txfonts}
\usepackage{verbatim} 
\usepackage{multirow,bigdelim} 
\usepackage{color}
\usepackage{vmargin}
\setmarginsrb{1.2cm}{1cm}{1.2cm}{1cm}{1cm}{1cm}{1cm}{1cm}

\newcommand{\hmsun}{{\, h^{-1}\rm~M}_\odot}

  \makeatletter
    \renewcommand{\paragraph}{\@startsection{paragraph}{4}{\z@}%
      {-3.25ex\@plus -1ex \@minus -.2ex}%
      {1.5ex \@plus .2ex}%
      {\normalfont\small\centering}}
     
    \renewcommand{\subparagraph}{\@startsection{subparagraph}{5}{\z@}%
      {-3.25ex\@plus -1ex \@minus -.2ex}%
      {1.5ex \@plus .2ex}%
      {\normalfont\small\centering}}
    \makeatother

\newcommand{\hMpc}{{ \textit{h}$^{-1}$~Mpc}}

\title[Fixed \& Paired]{Efficiently estimating mean, uncertainty and unconstrained large scale fraction of local Universe simulations with paired fixed fields}
\author[Sorce]
{{Jenny G. Sorce$^{1,2,3}$\thanks{E-mail: \text{jenny.sorce@ens-lyon.fr / jenny.sorce@univ-lyon1.fr / jsorce@aip.de}}, %
}\\
$^1$Univ Lyon, ENS de Lyon, Univ Lyon1, CNRS, Centre de Recherche Astrophysique de Lyon UMR5574, F-69007, Lyon, France\\
$^2$Univ Lyon, Univ Lyon1, ENS de Lyon, CNRS, Centre de Recherche Astrophysique de Lyon UMR5574, F-69230, Saint-Genis-Laval, France\\
$^3$Leibniz-Institut f\"{u}r Astrophysik (AIP), An der Sternwarte 16, D-14482 Potsdam, Germany\\
}

\begin{document}

\date{}

\pagerange{\pageref{firstpage}--\pageref{lastpage}} \pubyear{2020}

\maketitle

\label{firstpage}

\begin{abstract}

\indent 

Provided a random realization of the cosmological model, observations of our cosmic neighborhood now allow us to build simulations of the latter down to the non-linear threshold. The resulting local Universe models are thus accurate up to a given residual cosmic variance. Namely some regions and scales are apparently not constrained by the data and seem purely random. Drawing conclusions together with their uncertainties involves then statistics implying a considerable amount of computing time. By applying the constraining algorithm to paired fixed fields, this paper diverts the original techniques from their first use to efficiently disentangle and estimate uncertainties on local Universe simulations obtained with random fields. Paired fixed fields differ from random realizations in the sense that their Fourier mode amplitudes are fixed and they are exactly out of phase. Constrained paired fixed fields show that only 20\% of the power spectrum on large scales ($>$ tens of megaparsecs) is purely random. Namely 80\% of it is partly constrained by the large scale~/ small scale data correlations. Additionally, two realizations of our local environment obtained with paired fixed fields of the same pair constitute an excellent non-biased average or quasi-linear realization of the latter, namely the equivalent of hundreds of constrained simulations. The variance between these two realizations gives the uncertainty on the achievable local Universe simulations. These two simulations will permit enhancing faster our local cosmic web understanding thanks to a drastically reduced required computational time to appreciate its modeling limits and uncertainties.

\end{abstract}

\section{Introduction}
Within the past few years, reaching precision cosmology has been the driving force behind the tremendous efforts put in developing observational missions to acquire larger and larger cosmological large scale surveys \citep[e.g.][]{2006APS..APR.I7004B,2008dde..confE..33P,2019NatAs...3..574D} but also deeper local surveys \citep[e.g.][]{2001MNRAS.326..255C,2002MNRAS.334..673L,2004MNRAS.348.1355B} and in producing higher and higher resolution cosmological simulations with hydrodynamical physics \citep[e.g.][]{sch14,vogel14,schaye15,2016MNRAS.463.3948D}. However, if the standard or $\Lambda$CDM cosmological model reproduces very efficiently observations overall, tensions start to appear on various scales when pushing the comparisons down to the details \citep[e.g.][]{2016CQGra..33r4001S,2017ARA&A..55..343B,2017NatAs...1E.121F}, hence the redoubled efforts to confirm or infirm these apparent conflicts. \\

Clearly any systematic effect has to be dealt with \citep[see][for example of such systematics due to our local environment]{2010MNRAS.406...14F,2014MNRAS.438.1805W,2015JCAP...07..025W,2018ApJ...854...46H}. Measuring and understanding the impact of our local environment on our measurements and observations is one of the top priority. Knowing precisely our environment is, thus, an absolute prerequisite. That thirst for knowledge regarding our environment combined with recent increasing capabilities of numerical and observational technologies boosted considerably the studies of the local Universe. This renewed interest extended the definition of the local Volume to even greater distances so as to cover the full range of galaxy environments, from voids to massive groups and clusters. Nowadays, local is more and more commonly used for regions as large as about 300-400~Mpc \citep[e.g.][for a very few examples]{2013ApJ...775...62K,2012MNRAS.427L..35K,2016A&A...588A..14T,2016MNRAS.455.2078S,2018MNRAS.475.2519H}. \\

In this quest for the local Universe, cosmological simulations are combined with local observations in an attempt to achieve a fully complete picture of the local distribution of matter. This effort gave rise to the development of initial conditions constrained to result in simulations that resemble the local Universe at redshift zero. \\

Based on the concept introduced by Bertschinger in 1987 \citep{1987ApJ...323L.103B}, these initial conditions, in addition to abiding by a prior cosmological model like typical simulations \citep{1981MNRAS.194..503E} of large volumes \citep[e.g.][Euclid's flagship, for a non-exhaustive list]{2012MNRAS.426.2046A,DeusSimulation2012,2014MNRAS.444.1518V,2015MNRAS.446..521S,2016MNRAS.463.3948D,2016MNRAS.463.1797D}, also comply with a set of local observations, either densities obtained with redshift surveys \citep[e.g.][]{2006AJ....131.1163S,aih11,2011MNRAS.416.2840L,2012ApJS..199...26H} or peculiar velocities \citep[e.g.][]{1992ApJS...81..413M,1997ApJS..109..333W,2001MNRAS.326..375Z,2007ApJS..172..599S,2008ApJ...676..184T,2013AJ....146...86T,2016AJ....152...50T} or both. These constrained initial conditions are built either forwardly \citep[e.g][]{2008MNRAS.389..497K,2013MNRAS.432..894J,2013ApJ...772...63W}  or backwardly \citep[e.g.][]{1989ApJ...336L...5B,1990ApJ...364..349D,1999ApJ...520..413Z,1993ApJ...415L...5G,2008MNRAS.383.1292L}. Namely, the initial density field is either sampled from a probability distribution function (prior and likelihood given the observational data) or directly obtained from a realization of the density field today. These techniques resulted in multiple studies since then \citep[e.g.][for a very few examples]{2001ApJS..137....1B,2010MNRAS.406.1007L,2013MNRAS.429L..84K,2016MNRAS.455.2078S,2016MNRAS.457..172L,2016ApJ...831..164W,2016MNRAS.463.1462O,2017MNRAS.471.3087S,2017MNRAS.465.4886C,2018A&A...614A.102O,2018MNRAS.475.2519H,2018arXiv181111192O,2019MNRAS.486.3951S}. \\

However, these local Universe simulations present a common pitfall which is a residual cosmic variance, i.e. part of their properties is not constrained by the observational data but stays random. Any study requires then hundreds of runs before reaching sensible conclusions and their uncertainties. The latter are strongly linked to the residual cosmic variance or uncertainty on the local Universe simulations. Even in an era of expanding supercomputing facilities, decreasing the required computational time for a study is an appreciable advantage. Ideally one would want to get \emph{the} local Universe model but limitations due to the non-linearities of the problem, the limited size and resolution of the simulation box as well as the imperfect observational data makes it extremely challenging.  \\

Recent techniques diverted from their original use  provide an interesting alternative, to running hundreds of simulations, to evaluate efficiently uncertainties and to understand up to which level large scales local Universe simulations are constrained thanks to small scale~/ large scale correlations. Indeed, while previous studies focused on showing that the local Universe simulations either reproduce globally the large scales of the local Universe \citep[e.g][]{2009MNRAS.400..183K,2013MNRAS.432..894J,2014ApJ...794...94W,2016MNRAS.455.2078S,2016ApJ...831..164W} or the small scales down to the cluster scale in terms of mass and history \citep[e.g.][]{2016MNRAS.460.2015S,2018A&A...614A.102O,2018MNRAS.478.5199S,2019MNRAS.486.3951S}, inducing vaguely group scales \citep[e.g.][]{2016MNRAS.458..900C}, the fraction of large scales that can actually be  constrained - or reversely that cannot be constrained -  by a given dataset has never been evaluated.

On the one hand, \citet{2016MNRAS.462L...1A} proposed indeed cosmological simulations that dramatically decrease the sparse sampling of the largest wavemodes by fixing the initial Fourier mode amplitudes of the initial conditions \citep[e.g.][]{2010JCAP...06..015V}. Additionally, they paired them with a second set of initial conditions with initial modes exactly out of phase \citep{2016PhRvD..93j3519P}. They demonstrate that this technique drastically reduces the variance, namely their initial fields are not as random as typical initial fields anymore and their use in pairs permits deriving unbiased mean properties.

On the other hand, by definition constrained simulations reduce the cosmic variance with respect to typical simulations \citep[e.g.][]{2016MNRAS.455.2078S} but without suppressing it entirely, in particular in regions and on scales poorly constrained by the observational data. Our particular constraining algorithm relies on random realizations, in the sense that constraints from observational local data are applied to random fields to get plausible models of the local Universe. The residual cosmic variance and thus the uncertainty depends then on the random realizations used in the process. Consequently, it is difficult to estimate the uncertainty on a constrained simulation but for the runs of hundreds of other constrained simulations. It is also impossible to disentangle the randomness contribution from that of the small scale~/ large scale correlations to this uncertainty. It is thus challenging to gauge whether our current local data constrain anything at all on some scales and regions and further whether improvements are possible with upgraded local datasets. In the following, we include in the local dataset upgrades, an improvement of the technique used to constrain initial conditions. Thus any reference to an enhance local dataset refers to both a better catalog of constraints and a refine methodology to constrain.  \\

Instead of running hundreds of initial conditions obtained with random fields constrained with local observational data, this paper proposes to constrain ultimately two paired fixed fields of the same pair to understand and estimate uncertainties on local Universe simulations. The combination of the 'fixed-paired' method with the constraining algorithm allows us to 1) better understand the residual variance by splitting the contribution due to the large scale~/ small scale interactions from that induced by the randomization of the `unknown'  (fixing) and 2) obtain rapidly and efficiently an estimate of this residual variance and a mean realization of the local Universe by eliminating the influence of the random realization that is used (pairing).\\   

This paper opens on the definitions of the different types of fields and on a description of the methodology, of the different combinations of Gaussian (random), constrained, fixed and paired fields possible and thus of the simulations developed in this work. A second section focuses on the resulting simulations of interests for this paper goal. It compares them and highlights their properties before concluding. An appendix gathers all the possible combinations, including those out of focus for this paper, which could be used for further and other studies and goals.


\section{Constrained Gaussian (random), Fixed and Paired Initial Conditions}

\subsection{Gaussian (random), Fixed and Paired Fields}

We start with a short description of Gaussian (random), fixed and paired fields following the notations given by \citet{2018ApJ...867..137V} who gave a more exhaustive explanation. For a given density field, the overdensity (deviation of the density from the average at a given point) can be written $\delta(\vec k)=Ae^{i\theta}$ with $A$ and $\theta$ the amplitude and the phase of the mode $\vec k$. Subsequently, 
\begin{itemize}
\item \emph{for a Gaussian field}, $\theta$ is a random variable uniformly distributed between 0 and 2$\pi$ and $A$ follows a Rayleigh distribution:
\begin{equation}
p(A)dA=\frac{A}{(VP(k)/16\pi^3)^2}e^{-A^2/2(VP(k)/16\pi^3)^2}dA
\end{equation}
where $P$ is the power spectrum and V the volume of the simulation box.\\
\item \emph{for paired Gaussian fields}, the second Gaussian field of the pair is out-of-phase by a factor $\pi$ namely $\theta$ becomes $\theta + \pi$ or it is simply the opposite of the first one: $\delta(\vec k)=Ae^{i\theta+\pi}=-Ae^{i\theta}$.\\
\item \emph{for a fixed field}, while $\theta$ is unchanged, $A$ follows a distribution with identical values as in the Gaussian fields, i.e. $\langle\delta(\vec k)\delta^*(\vec k)\rangle=VP(k)/(2\pi)^3$,  but without intrinsic scatter:
\begin{equation}
p(A)dA=\delta^D\left(A-\sqrt{\frac{VP(k)}{(2\pi)^3}}\right)dA
\end{equation} 
\item \emph{for paired fixed fields}, the second fixed field is out-of-phase by a factor $\pi$.
\end{itemize}

\subsection{Constrained Gaussian (random) Fields}

A detailed description of the technique to build constrained initial conditions from Gaussian (random) fields is outside the scope of this paper and we refer the reader to \citet{2016MNRAS.455.2078S} and \citet{2018MNRAS.478.5199S} for more explanations. A brief description using galaxy radial peculiar velocities as constraints is as follows:
\begin{enumerate}
\item Grouping of the radial peculiar velocity catalog to remove non-linear virial motions that would affect the linear reconstruction obtained with a linear method \citep[e.g.][]{2017MNRAS.468.1812S,2018MNRAS.476.4362S}. More precisely, when distance estimates are available for several galaxies in a given cluster, these estimates are replaced by the distance of the cluster. The cluster peculiar velocity is then derived. This one constraint is used for the cluster.
\item Minimizing the biases \citep{2015MNRAS.450.2644S} in the grouped radial peculiar velocity catalog and attributing the residual uncertainties \citep{2018MNRAS.478.5199S}. Biases are indeed inherent to any observational catalog. An iterative technique, based on the sole prior that the distribution of radial peculiar velocity should be a Gaussian with a variance determined by the cosmological model, can drastically reduce biases. Residual uncertainties can then be estimated depending on the distance to the observer and the amount of data as a function of this distance.
\item Reconstructing the 3D cosmic displacement field with a linear minimum variance estimator or Wiener Filter, \citep[in abridged form WF,][]{1995ApJ...449..446Z,1999ApJ...520..413Z} applied to the radial peculiar velocity constraints.
\item Relocating constraints (galaxies and their velocities) at the positions of their progenitors using the Reverse Zel'dovich Approximation and the reconstructed cosmic displacement field \citep{2013MNRAS.430..888D} and replacing noisy radial peculiar velocities by their 3D WF reconstruction \citep{2014MNRAS.437.3586S}. Subsequently, one can expect structures to be at the proper position, i.e. at positions similar to those observed, at the end of the simulation run.
\item Producing density fields constrained by the modified observational peculiar velocities combined with a Gaussian (hereafter random) realization to restore statistically the ``missing'' structures (the WF goes to the null field in absence of data or in presence of very noisy data). It means obtaining an estimate of the residual between the model and the data. The Constrained Realization technique \citep[][]{1991ApJ...380L...5H,1992ApJ...384..448H}, which differs schematically from the WF by a random realization added to the constraints, is used for that step.
\end{enumerate}

\subsection{Constrained, Fixed and Paired Fields}

Given the algorithms to build the different fields (fixed, paired, constrained), several types of initial conditions can be produced by \#1 changing the realization in step (v) described above and possibly \#2 adding one further step (vi) when building the initial conditions. 

\begin{figure}
\centering
\includegraphics[width=0.48 \textwidth]{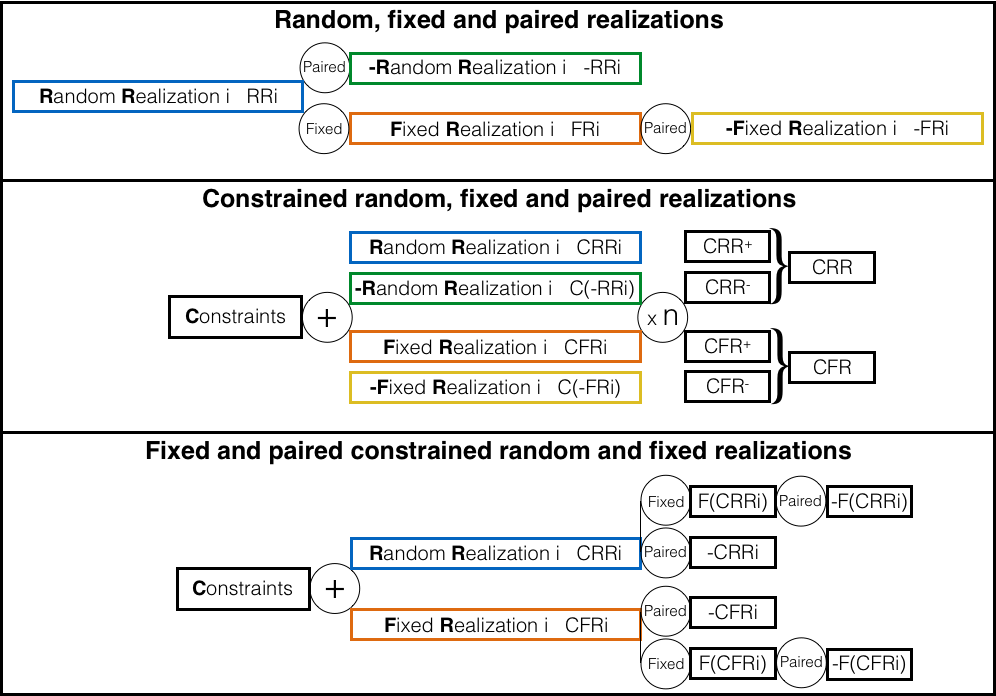}
\caption{This figure summarizes the different types of fields that can be constrained (top), the different types of constrained fields (middle) and the additional steps that can be applied to constrained fields (bottom). i stands for the seed used to obtain the field. n is the number of different seeds or fields, n=50 at maximum in this paper.}
\label{fig:tbl}
\end{figure}

The first panel of Figure \ref{fig:tbl} gives the different types of fields that can be constrained in step (v) with local observational data as well as their connection to each other:\vspace{-0.25cm}
\begin{itemize}
\item Gaussian or Random Realization (RR)
\item Fixed Realization (FR)
\item Paired Random Realization (-RR)
\item Paired Fixed Realization (-FR)
\end{itemize}
\#1 consists then in selecting any of these fields for step (v) of the initial conditions building process to get initial conditions as enumerated in the second panel of Figure \ref{fig:tbl}.
\#2  proposes after producing the initial conditions to additionally fix and/or pair them (step (vi)) as listed in the last panel of Figure \ref{fig:tbl}.\\

\begin{table}
\begin{center} 
\begin{tabular}{cccccc}
\hline
\hline
& Constrained & Random & Fixed & Paired & Seed   \\
& & Realization & Realization &&\\
 & & & & &\\
Symbol & C & RR & FR & -  & `Number'\\
\hline
 \multicolumn{6}{c}{Combinations in this paper, abbreviations}\\
\multicolumn{2}{r}{ Type of fields} & \textbf{Random} & \textbf{Paired} & \textbf{Fixed} & \textbf{Paired} \\
 \multicolumn{2}{l}{Type of }       &  & \textbf{Random} & &  \textbf{Fixed} \\
 \multicolumn{2}{l}{constrained fields} &       & &&\\
\multicolumn{2}{l}{ \textbf{Constrained}} & \textbf{CRRi}  & \textbf{C-(RRi)}    & \textbf{CFRi}  &\textbf{C-(FRi)}   \\
 \multicolumn{2}{l}{Paired Constrained} &  -CRRi        &    &    -CFRi      & \\
 \multicolumn{2}{l}{Fixed Constrained} & F(CRRi) & & F(CFRi) &  \\
  \multicolumn{2}{l}{Paired Fixed Constrained} & -F(CRRi)  &  & -F(CFRi) \\
\hline
 \multicolumn{6}{c}{Other notations}\\
 \multicolumn{6}{l}{Hereafter unpaired sets:} \\ 
 \multicolumn{5}{l}{Set of n CRRi} & CRR$^+$\\
 \multicolumn{5}{l}{Set of n C(-RRi)} & CRR$^-$\\
 \multicolumn{5}{l}{Set of n CFRi} & CFR$^+$\\
 \multicolumn{5}{l}{Set of n C(-FRi) } & CFR$^-$\\
  \multicolumn{6}{l}{}\\ 
  \multicolumn{6}{l}{Hereafter paired sets:} \\ 
  \multicolumn{5}{l}{Set of n CRRi and n C(-RRi)} & CRR$\textcolor{white}{^+}$\\
    \multicolumn{5}{l}{Set of n CFRi and n C(-FRi)} & CFR$\textcolor{white}{^+}$\\
 \hline
\hline
\end{tabular}
\end{center}
\vspace{-0.25cm}
\caption{Symbol and given short names to the different simulations as well as sets of simulations. In the second block, bold letters stand for simulations studied in details, others are summarized in the appendix. While, i represents the seed used to obtain the field, n (=50 at maximum, here) is the number of different seeds.}
\label{Tbl:1}
\end{table}

Our ultimate goal would be to reproduce the local Universe as precisely as possible. Ideally, the residual variance between the different constrained fields should approach zero so as to get the local Universe initial conditions and with its numerical evolution, the local Universe at z=0. Given the challenge and current limitations, our second approach is to understand and fully estimate the uncertainty on local Universe simulations. Given this goal, the combinations of interests for this paper are those giving the constrained (paired) random and fixed fields obtained with \#1. Fields that are fixed and/or paired after constraining (\#2) will not retain the intermediate scale structure of the local Universe: namely clusters, especially the `Centaurus - Virgo pair', with proper masses have less probabilities of forming in a posteriori fixed fields since the probability of this `cluster pair' is not high. A fixed field gives the most common structures. As for paired constrained fields, they produce voids at local cluster positions and vice versa. These fields are, thus, not of immediate interests for this paper goal.

Consequently, although Table \ref{Tbl:1} enumerates the short names given to all these simulations, the next section focuses on the constrained (paired) random and fixed fields, while paired and/or fixed constrained fields are relegated to the sole appendix. This appendix summarizes the different types of simulations and gives their features to emphasize the difference between modifying step (v) or adding step (vi).\\

The next section shows the benefits of constraining fixed realizations as well as constraining a pair of paired realizations. These simulations differ from typical constrained simulations in the sense that the constraints are applied to specific paired and fixed realizations rather than random realizations to obtain their initial conditions. They do not pretend to result in the best local Universe initial conditions, but rather will permit \#1 disentangling a purely random component from a partly constrained component to the residual cosmic variance on large scales and \#2 fully estimating this residual cosmic variance or uncertainty on local Universe simulations thus on conclusions drawn from them.


\section{Constrained Paired Fixed and Random fields}

\subsection{Power spectrum \& density field}

\begin{figure}
\centering
\includegraphics[width=0.45 \textwidth]{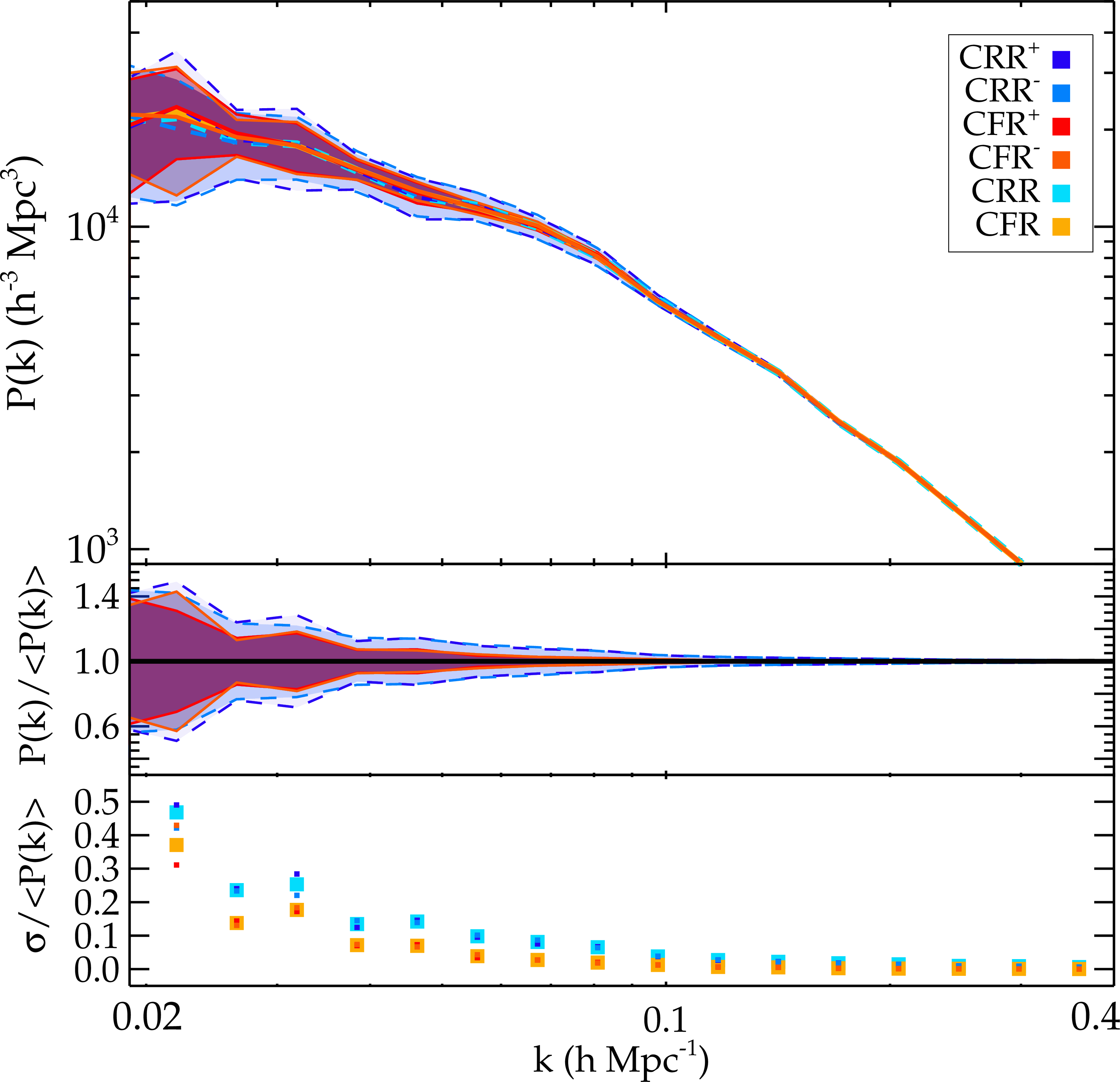}
\caption{Top: Power spectrum of constrained initial conditions. Their standard deviations are shown as red and orange (dark and light blue) transparent areas delimited by thin solid (dashed) lines of the same color for unpaired sets of fixed (random) fields.  Their means are given by thick solid and dashed lines of the same color for CFR$^{\pm}$ and CRR$^{\pm}$ respectively. Additionally, the mean power spectrum of the entire sample or paired set of constrained fixed fields, CFR (constrained random fields, CRR), is shown as a thick yellow (cyan) solid (dashed) line.   Middle: Power spectra divided by their respective mean, same color code. Bottom: Variance of the power spectra divided by their respective mean, same color code.}
\label{fig:powspecics}
\end{figure}

50 constrained initial conditions (100 in total after pairing) were prepared with n=50 different random realizations (n $\times$ RRi) and n=50 others (100 in total after pairing) were obtained with 50 different fixed realizations (n $\times$ FRi). All the initial conditions were built at redshift 60 using 256$^3$ dark matter particles (particle mass 6.4$\times$10$^{11}\hmsun$) in 500~\hMpc\ boxes within the Planck cosmology framework \citep[$\Omega_m$=0.307, $\Omega_\Lambda$=0.693, H$_0$=67.77~km~s$^{-1}$~Mpc$^{-1}$, $\sigma_8$~=~0.829,][]{2014A&A...571A..16P},\\

Figure~\ref{fig:powspecics} presents the power spectrum of the sets of constrained initial conditions obtained with 50 random realizations (CRR$^+$), 50 fixed realizations (CFR$^+$), 50 paired random realizations (CRR$^-$) and 50 paired fixed realizations (CFR$^-$). For the largest modes present in the box, the variance with respect to the mean (thick dashed and solid lines) is overall smaller by 20\% for the constrained initial conditions obtained with the FRi (orange and red areas) than with the RRi (light and dark blue areas). This is half expected since the fixed field without constraints have the exact same power spectrum values. Adding the local constraints does not re-introduce the full residual cosmic variance. 

However, it confirms that almost 80\% of the scatter is due to correlations between the local constraints (a few megaparsecs) and the large scales (tens of megaparsecs). Namely, the residual cosmic variance, far from being due to the sole random realization, is mostly due to the correlations between the constraints and the realization. Thus these scales are partly constrained. It means that, for a given constrained random realization of the local Universe, 80\% of the scatter could be further reduced by enhancing the local dataset used as constraints within the 500~\hMpc\ boxsize. 20\% of the power spectrum on larges scales though is not constrained. Larger boxsizes would thus later be required to continue diminishing the power spectrum residual cosmic variance to get \emph{the} local Universe model even at the 50~\hMpc\ scales. \\

Additionally, it is interesting to notice that the entire or paired sets of initial conditions (i.e. obtained either with n$\times$RRi and n$\times$-(RRi) or n$\times$FRi and n$\times$-(FRi)) are better representative of the mean and variance with respect to the mean for the largest modes than the unpaired sets (i.e. obtained alternatively with n$\times$RRi,  n$\times$-(RRi), n$\times$FRi or n$\times$-(FRi)). It should also be pointed that means obtained with either the constrained paired fixed fields or the constrained paired random fields are remarkably similar. Namely, no bias is introduced by using fixed rather than random fields as a basis to produce constrained initial conditions.

\begin{figure}
\centering
\includegraphics[width=0.47 \textwidth]{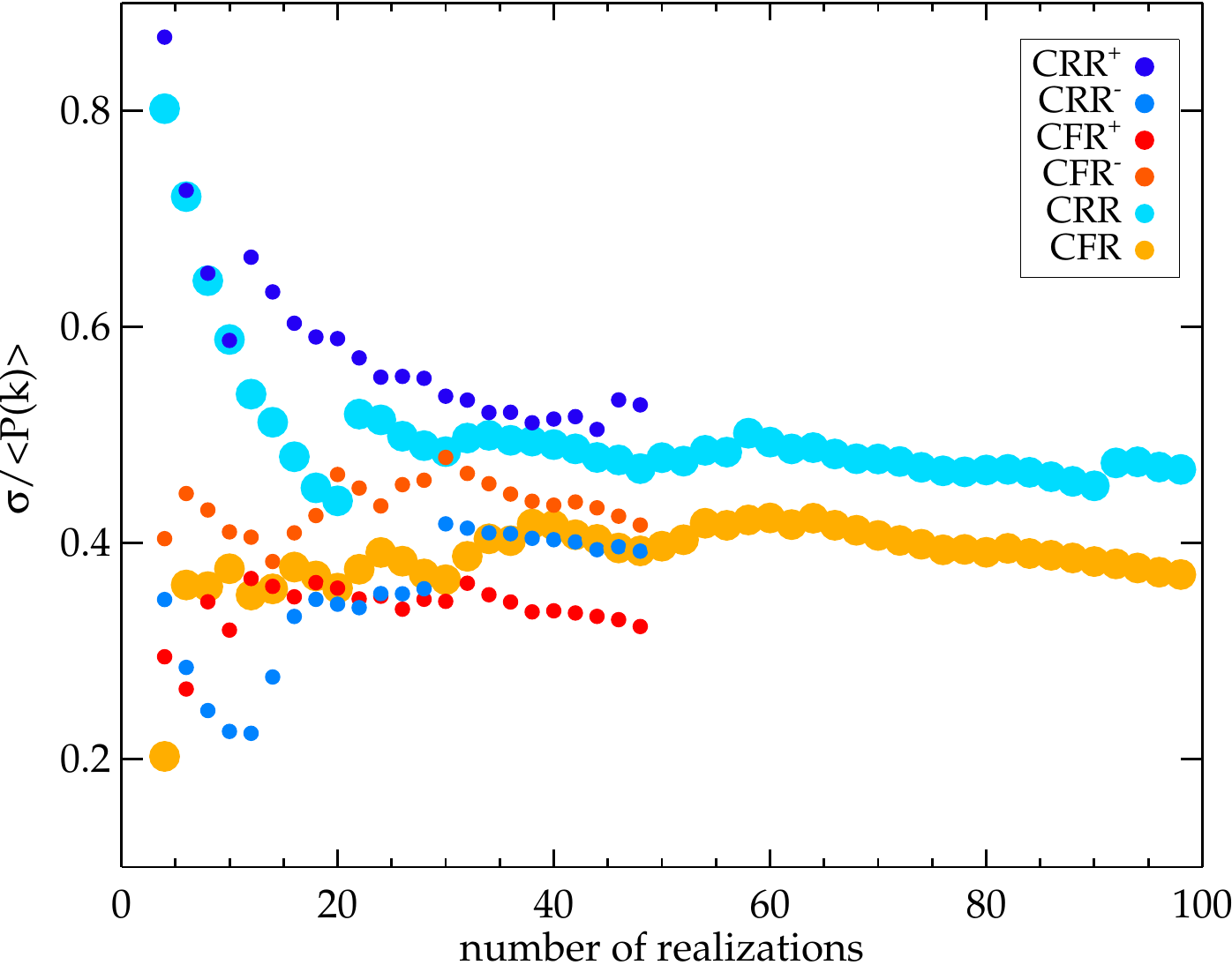}
\caption{Variance of the power spectra divided by their mean for the largest scale mode valid in the box as a function of the number of constrained realizations in the set of initial conditions used to derive the variance. Constrained fixed realizations (CFR$^\pm$) are shown as red and orange filled circles. Constrained random realizations (CRR$^\pm$) are represented by light and dark blue filled circles. The yellow and cyan larger filled circles are obtained using the same number of realizations as in the unpaired sets but for the paired sets (CRR and CFR).}
\label{fig:sumuppow}
\end{figure}

However, it is clear that the paired sets are required to get unbiased results and means in both cases. For instance, in the bottom panel of the figure, only the yellow and cyan filled squares, which stand for the ratio of the power spectrum variance to the power spectrum mean, clearly show that the scatter is smaller for the constrained initial conditions obtained with both the FRi and -FRi than for those obtained with both the RRi and -RRi. On the opposite, half sets, obtained with only the -RRi and -FRi, show a reversed result for the largest scale mode valid in the box or more precisely no clear difference. Tests with a smaller number of realizations per unpaired set reveal that this is even truer the smaller this number is. \\

 Actually, Figure \ref{fig:sumuppow} shows the ratio of the power spectrum variance to the mean for the largest scale mode valid in the box as a function of the number of realizations included in the unpaired and paired sets of constrained initial conditions. As shown with the yellow and cyan filled circles compared to the smaller blue and orange/red filled circles, clearly the mean variance is reached faster when pairing. Again fixing permits understanding that about 80\% of the residual cosmic variance is due to correlations between the small scales observational constraints and the large scales. It is also clear that pairing is absolutely necessary to reach the proper conclusion when using a small set of simulations to derive the variance: blue and red/orange filled circles show biased values.\\
 
Both profits of applying the constraints to fixed and paired fields to understand and estimate uncertainties on local Universe simulations appear already and can be summarized as follows : \\
$\bullet$ by reducing the residual cosmic variance of the power spectrum of local Universe initial conditions by 20\% for the largest modes in the box, fixing shows that about 80\% of the power spectrum on large scales is partly constrained. It is thus not completely random. It implies that enhanced datasets are already useful to decrease some more the residual cosmic variance on scales as large as 50~\hMpc\ before thinking about enlarging the boxsize of the simulations. \\
$\bullet$ pairing permits recovering the residual variance of the power spectrum between the different realizations more efficiently, namely quicker, and results in unbiased values : a smaller set of initial conditions is required. We will also show later that a combination of constrained paired fields of the same pair permits recovering efficiently the mean of several constrained fields. It thus gives instantaneously the residual variance or uncertainty on the simulated local Universe in an unbiased way.\\

All the prepared initial conditions (50 per set, 100 in total after pairing) are run from redshift 60 to redshift 0 using gadget \citep{Springel2005}.\\

Power spectra are derived with the density fields obtained from a cloud-in-cell scheme applied to the simulation snapshots at redshift zero. Results for the power spectrum of the evolved initial conditions are similar to those obtained for the power spectrum of the initial conditions shown in Figure \ref{fig:powspecics}. The exact same conclusions can thus be reached. \\

\begin{figure*}
\includegraphics[width=1 \textwidth]{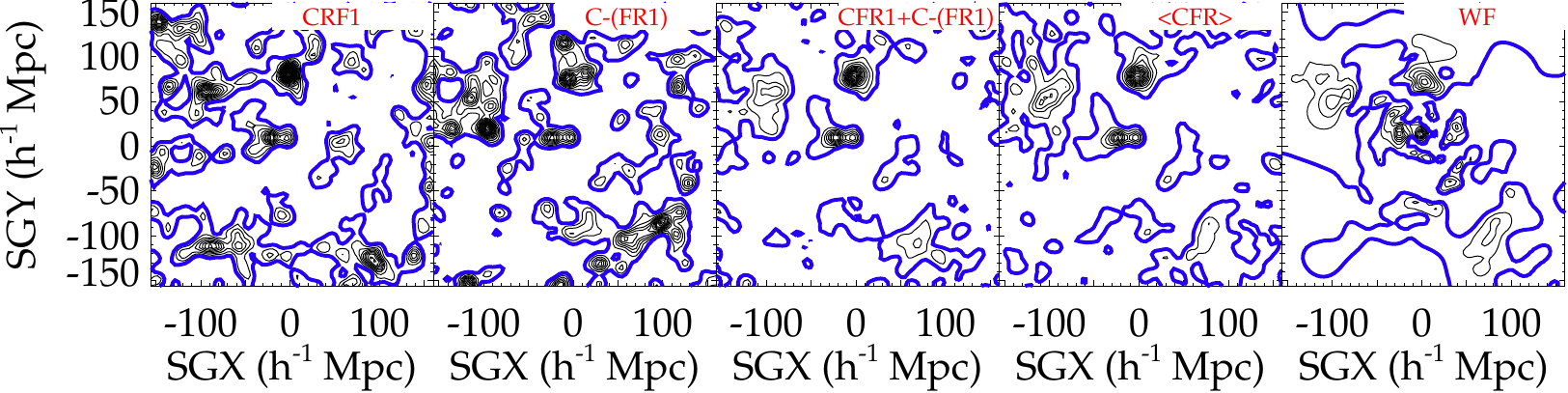}
\caption{XY supergalactic 2~\hMpc\ thick slice of the density fields, smoothed at 5~\hMpc\ of the simulations/reconstruction at redshift zero of the local Universe. Black contours show the overdensities while the solid blue lines represent the mean density. From left to right: simulation obtained with a constrained fixed field, simulation based on the constrained paired fixed field, geometric mean of the two constrained paired fixed fields of the same pair, geometric mean of all the simulations obtained with fixed paired fields, reconstruction obtained with the Wiener filter technique.}
\label{fig:density}
\end{figure*}

From left to right, Figure \ref{fig:density} shows the XY supergalactic 2~\hMpc\ thick slice, of the smoothed at 5~\hMpc\ density field of 1) and 2) a pair of constrained paired fixed fields, 3) their geometric mean as well as 4) that of all the different constrained paired fixed fields and 5) the Wiener filter reconstruction. Black contours stand for the overdensities in the slice and the blue color delimits the overdensities from the underdensities. In the first two panels, the constrained paired fixed density fields show that the local large scale structure is recovered: Virgo, Centaurus (both are close to the center of the slice) and Coma (at about Y=70-80~\hMpc) regions are visible. The Shapley region is also overdense (XY$\sim$[-100, 50]\hMpc). \\

Because the density properties of paired fields are expected to be almost the opposite, namely where there are peaks in the field, voids are expected in the paired field although not necessarily of the same magnitude \citep[see][for a detailed explanation]{2016PhRvD..93j3519P}, it is interesting to combine constrained paired fields of the same pair to estimate the `constraining power' of the constraints. In other words, combining constrained paired fields of the same pair allows determining the features of the local Universe that are robustly simulated and at which level. A part of the field that is not constrained should thus be annihilated when combined with its counterpart in the paired field. Reversely, a structure that is solidly constrained should persist after taking the geometric mean of the constrained paired fields of the same pair. A persistence gradient should exist for structures partly constrained. This gradient gives the uncertainty on the simulated structure: there is no uncertainty on a structure 100\% persistent, it exists in the local Universe and it is robustly simulated. On the opposite, a structure that disappears is uncertain, perhaps does not exist in the local Universe and in any case is poorly constrained. 

From our previous studies, we expect Virgo \citep[][]{2016MNRAS.460.2015S,2019MNRAS.486.3951S} to be very well constrained, then comes Centaurus, then Coma, etc with a decreasing `constraining power' with the distance from us \citep{2016MNRAS.455.2078S}. The fourth panel of Figure \ref{fig:density} shows the geometric mean of all the constrained paired fixed fields. As expected, Centaurus, Virgo and Coma regions appear very well constrained, then comes the Shapley region and to a lesser extent the Perseus Pisces region. This region is known not to be well constrained yet because of the weak amount of data in this region in the catalog of constraints used so far \citep{2013AJ....146...86T}.\\

Interestingly, the geometric mean of only two constrained paired fixed fields of the same pair in the third panel is very similar to the geometric mean of all the constrained simulations. Two constrained simulations obtained with paired fixed fields of the same pair are thus qualitatively capable of reproducing the mean of an ensemble of constrained simulations. The last panel shows the linear reconstruction of the local Universe as a sanity check of the non-linear simulations of the local Universe.\\

Note that the figure obtained when constraining paired random realizations rather than paired fixed realizations is similar to Figure \ref{fig:density}. The same conclusions can thus be reached. \\

Eventually, while fixing gives us a better understanding of the fraction of the residual cosmic variance that can be better constrained by local enhanced data, pairing allows us to estimate the residual cosmic variance very efficiently. It is given by the standard deviation between the two constrained fields of the same pair. It corresponds to the uncertainties on the simulated structures, in other words its gives the confidence in the simulation to reproduce the local Universe.

\begin{figure*}
\vspace{1cm}
\begin{minipage}[t]{.48\textwidth}
\vspace{-8cm}
\includegraphics[width=1 \textwidth]{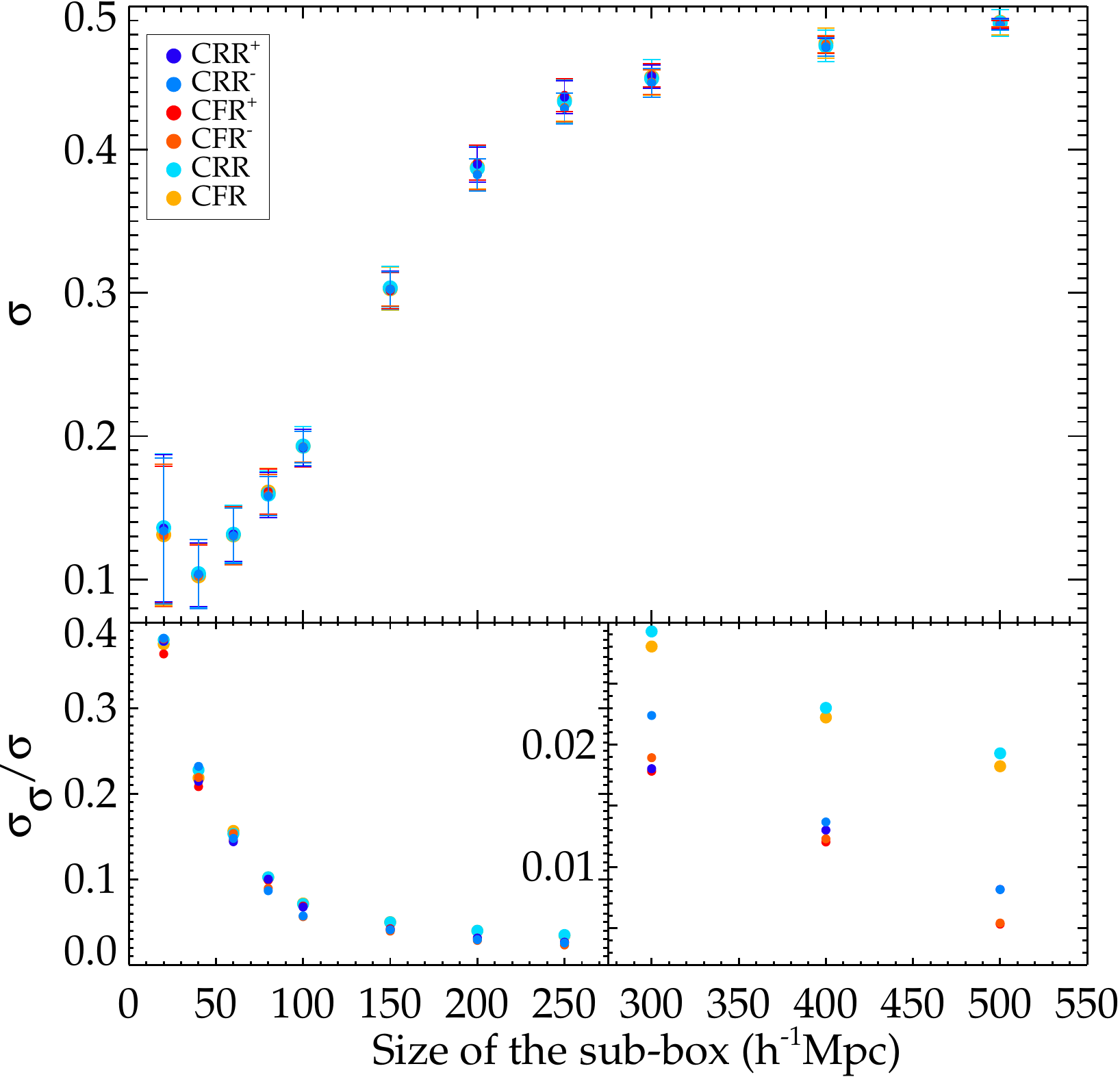}
\end{minipage}
\begin{minipage}[t]{.48\textwidth}
\includegraphics[width=1 \textwidth]{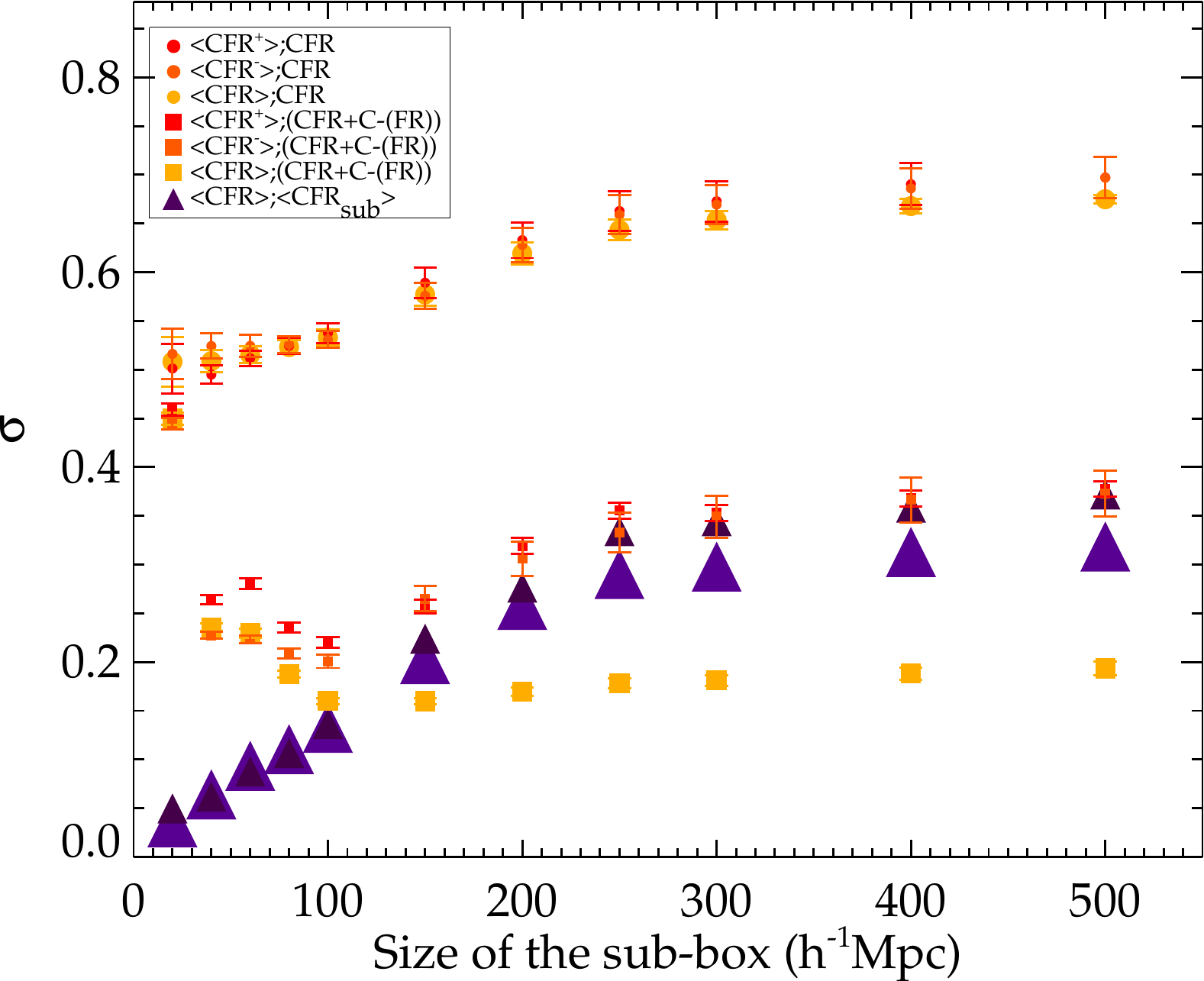}
\end{minipage}

\caption{Top left and right: Average residual cosmic variance (filled symbol) and its standard deviation (error bar) between density fields of simulations (left) or between density fields of simulations and their geometric means (right) as a function of the size of the compared sub-box. See Table \ref{Tbl:1} for an explanation regarding the abbreviations. Bottom left: standard deviation of the residual cosmic variance to the residual cosmic variance ratios as a function of the size of the compared sub-box, same color code. The geometric mean of two constrained paired fixed simulations of the same pair is as good a proxy of the geometric mean of an (independent) set of constrained fixed simulations (right panel) as, if not a better one than, an ensemble of independent constrained fixed simulations (filled triangles with size proportional to the number of simulations in the set).}
\label{fig:densityvar}
\end{figure*}

Additionally, pairing gives a fast estimate of what the mean of hundreds of local Universe simulations would be. In order to check that the value of the mean is unbiased when using the fixing process with respect to not using it, we compare the geometric mean of the two constrained paired fixed fields of the same pair to that of the two constrained paired random fields of the same pair. The geometric means share the same mean (-0.1), standard deviation (0.3), maximum (8) and minimum (-0.8) density values. Their mean difference is of the order 10$^{-4}$ with a standard of 0.16 (Figure \ref{fig:densityvar} right, below, shows that this value is smaller than the difference between the geometric mean of two constrained paired fields and that of all the constrained fields). Numbers are given in units of density. There is thus no bias.\\

It is interesting to quantitatively derive~:
\begin{itemize}
\item \#1 an estimate of the residual cosmic variance or mean variance between different constrained simulations as well as the standard deviation or uncertainty of this residual cosmic variance.
\item \#2 the ability of two constrained paired (fixed) fields of the same pair to reproduce the mean of several constrained simulations. Typically it gives the existence certainty of structures, in other words the `constraining power' of the constraints used.
\end{itemize}

To that end, cell-to-cell comparisons between pairs of simulations are conducted. First, cells are compared within the full box. The scatter around the 1:1 relation is derived. Once all the scatters are obtained for a given type of simulation pairs, their mean and variance are computed. Second, because simulations are known to be more constrained in the center of the box where most of the constraints are, cells are compared only in sub-boxes. All the resulting mean scatters (as defined above when comparing the full boxes and different size sub-boxes) and their variances are reported in Figure \ref{fig:densityvar} left as a function of the size of the sub-boxes within which cells are compared between simulations. \\

\#1 The trend is perfectly similar whatever set of constrained simulations is considered. As expected, the residual cosmic variance between the different density fields is the smallest in the inner part of the box where most of the constraints are \citep{2016MNRAS.455.2078S}. At this level there is no difference between residual cosmic variances obtained constraining the fixed fields rather than the random ones. Every inch of the density field is somewhat constrained. It confirms our previous findings that cosmic variance is reduced even when considering the entirety of the 500~\hMpc\ box although constraints are restricted to the inner $\sim$300~\hMpc\ \citep{2016MNRAS.455.2078S}.

Considering the variance of this residual cosmic variance or in other words, the uncertainty on the residual cosmic variance, we expect to find the largest variance for pairs involving paired fields. Indeed, the bottom panels of Figure \ref{fig:densityvar} left show that the largest values are obtained for the set of simulations including both constrained paired fields (yellow and cyan filled circles against red, orange, light and dark blue filled circles). The values are almost doubled when considering the full boxes. This reinforces our claim that pairing allows determining the full residual cosmic variance in an unbiased way.

In addition, there is a slight hint that simulations obtained with the fixed fields present a slightly smaller variance of the residual cosmic variance to the latter ratio than those obtained with the random fields. The decrease is less than a few percent though. It is thus legitimate to consider that at the density level, an unbiased residual cosmic variance can be independently derived with constrained fixed fields or constrained random fields as long as they are properly paired.\\

\# 2 Figure \ref{fig:densityvar} right gives a quantitative measurement of the ability of the geometric mean of the same pair constrained paired fixed fields to reproduce efficiently the geometric mean of several constrained fixed fields. The plot is similar for non-fixed fields. As before, cell-to-cell comparisons between pairs of density fields are conducted in different sub-boxes. Density fields that are used for comparisons are 1) the geometric mean of an ensemble of paired and unpaired constrained fixed fields and 2) the single or combined constrained paired fixed fields of the same pair. The variance is significantly smaller between the geometric mean of the same pair constrained paired fixed fields and that of all the constrained fixed fields than between a single fixed field and the geometric mean of all the fields. Whatever sub-box size and geometric mean (that of CFR$^+$, CFR$^-$, CFR) are used for comparisons, it is only about 20\% (yellow filled squares) in the former case against about 3 times more (about 60\%, filled circles) in the latter case. 

Comparing the variance between the geometric mean of all the constrained fields to 1) that of the same pair constrained paired fixed fields (yellow filled squares) and 2) to that of several independent constrained fixed fields (filled triangles) shows that overall the same pair constrained paired fixed fields are better representative of the mean of all the fields than 50 independent constrained fixed fields (large violet filled triangles). Moreover, 25 independent constrained fixed fields (small dark violet filled triangles) are required to reach a variance as low as that obtained when comparing the geometric mean of the same pair constrained paired fixed fields to the geometric mean of all the independent (i.e. not paired) constrained fields.  This observation is in favor of our claim that the geometric mean of two constrained paired fixed fields of the same pair is a good proxy for the geometric mean of an ensemble of constrained fixed fields be they independent or not. 

In other words, in a first approximation two simulations are completely sufficient to determine the structures of the local Universe that are actually constrained (exist) and up to which level of confidence in an unbiased way. This second part is linked to the fact that the largest variance is obtained when comparing constrained paired fields of the same pair as shown on the left part of Figure \ref{fig:densityvar}. Namely, the full residual cosmic variance or the full uncertainty estimate on the simulated structures is given by the variance between the constrained paired fields of the same pair. This variance is a good non-biased proxy for the uncertainty on the simulated structures.\\

It is to be noted that this conclusion seems untrue for small sub-box sizes since a reversed trend is visible: the geometric mean of the same pair constrained paired fields does not reproduce the mean as well as a set of independent and randomly selected constrained fields. This is not unexpected. The larger the number of fields used to derive the mean the more the small scale details are erased. Small shifts of structures start then to have higher effects on cell-to-cell comparisons when one tries to maintain a sufficient number of cells for statistical comparisons. This observation is thus a clear limit of the method used to quantify the difference between the density fields rather than a flaw in our conclusion. \\ 

\begin{figure*}
  \includegraphics[width=0.48 \textwidth]{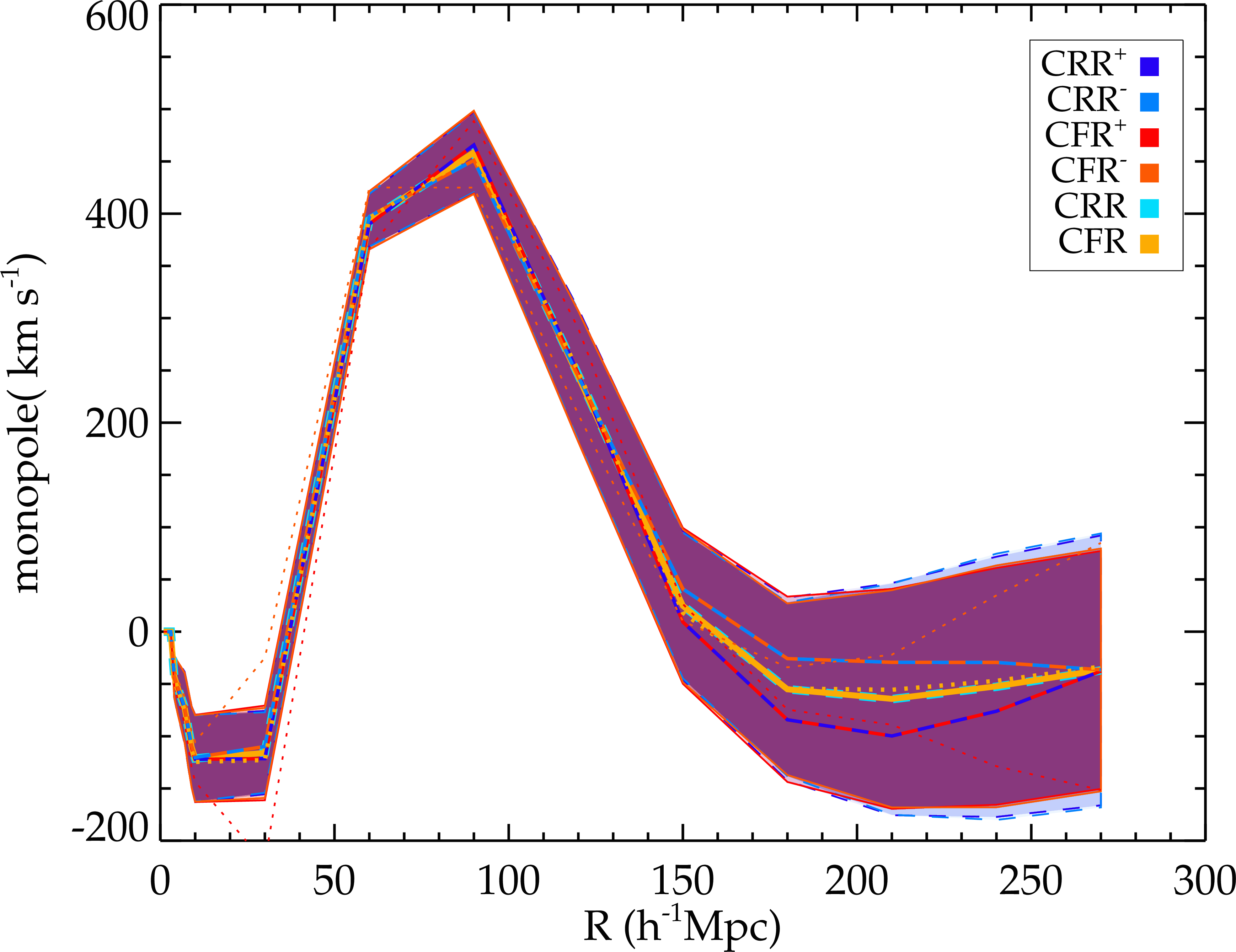}
  \includegraphics[width=0.48 \textwidth]{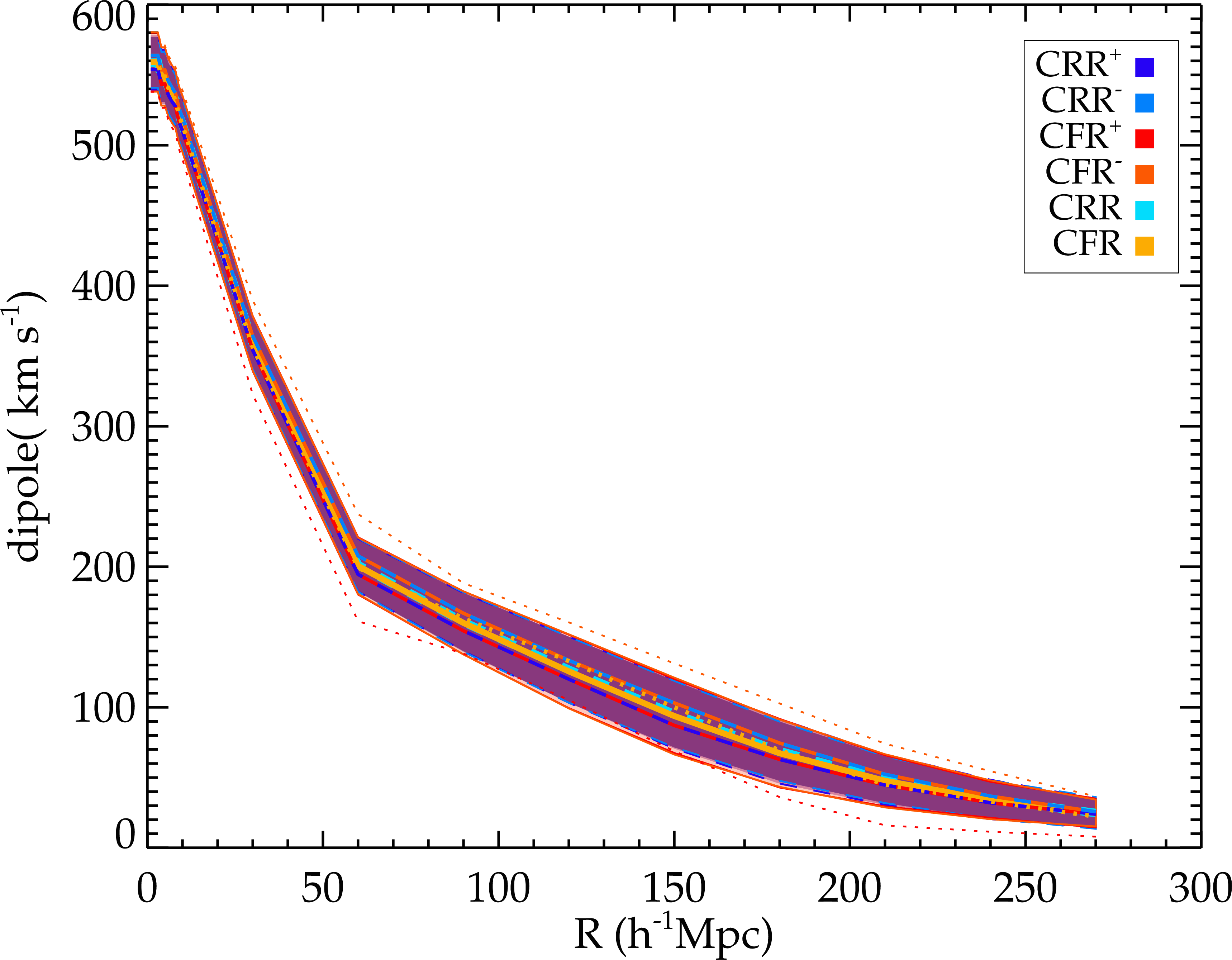}
\caption{Monopole (left) and dipole (right) of the velocity fields of the constrained simulations. Same color code as Figure \ref{fig:powspecics}. The dotted lines give an example of the values for two constrained realizations of the same pair (red and orange) as well as their mean (yellow).}
\label{fig:monodip}
\end{figure*}

Figure \ref{fig:monodip} pursues further the investigation with the example of the monopole (left) and the dipole (right) of the velocity fields. First it confirms that using fixed rather than random realizations does not bias our results. Means (solid and dashed lines) and scatter (transparent areas) of monopoles and dipoles are identical when comparing sets of constrained fixed and random fields of the same nature (CFR / CRR, CFR$^{\pm}$~/~CRR$^{\pm}$). Additionally, the dotted lines highlight again that combining a pair of constrained paired fixed realizations (the mean of the red and orange dotted lines gives the yellow dotted line) permits retrieving immediately the mean of several constrained realizations, since the yellow solid and dotted lines are similar. Although this is not shown to preserve the readability of the figure, the same conclusions are valid for a pair of constrained paired random realizations.
 
 \subsection{Mass functions \& Halos}
 
\begin{figure}
\centering
\includegraphics[width=0.45\textwidth]{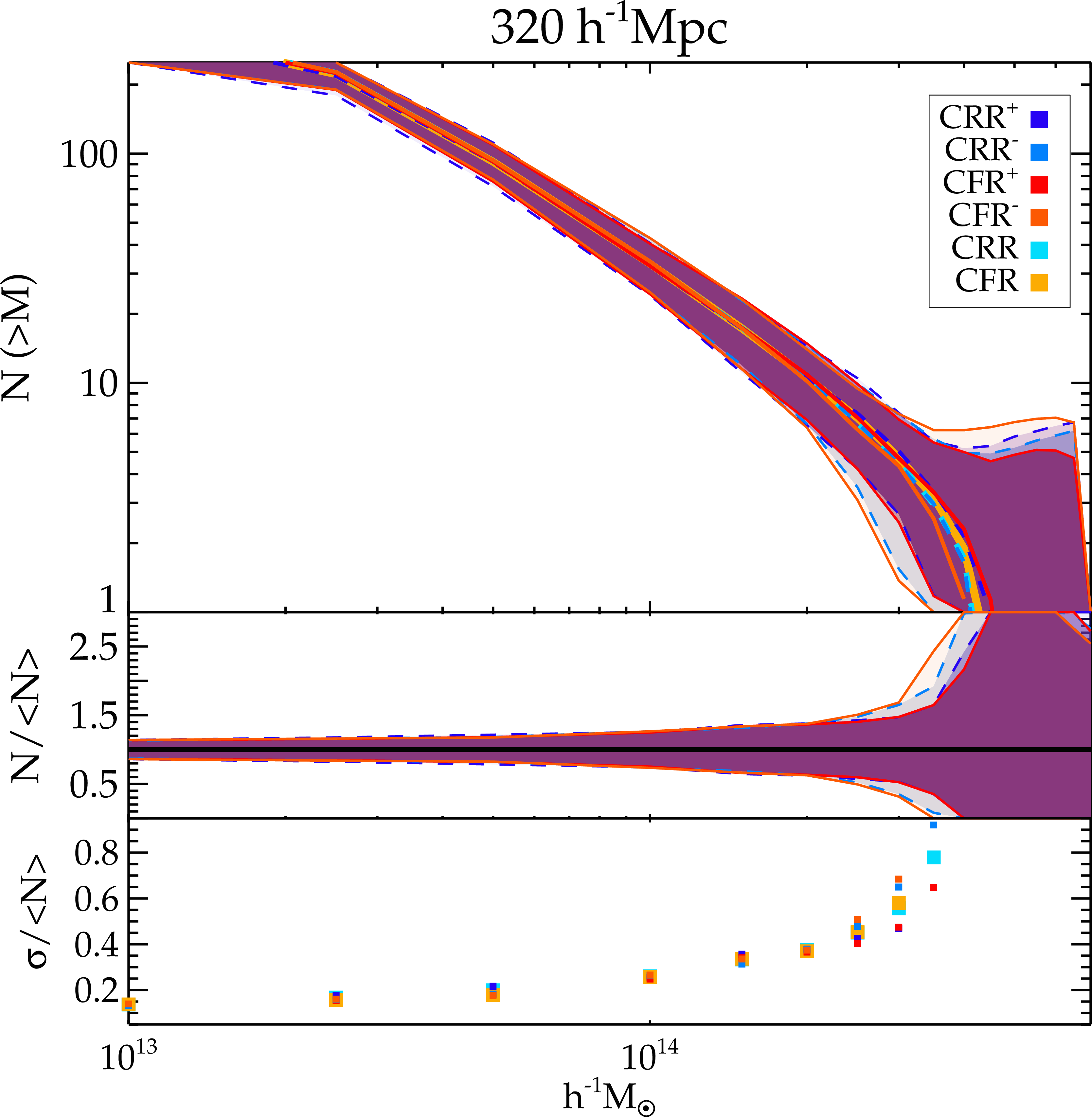}\\
\caption{Top: Mass functions of a 160~\hMpc\ radius sphere centered on the box middle of constrained simulations. Their standard deviations are shown as red and orange (dark and light blue) transparent areas delimited by thin solid (dashed) lines of the same color for paired fixed (random) fields.Their means are given by thick solid and dashed lines of the same color for CFR$^{\pm}$ and CRR$^{\pm}$ respectively. Additionally, the mean mass function of the entire sample of constrained fixed fields, CFR (constrained random fields CRR), is shown as a thick yellow (cyan) solid (dashed) line. Middle: Ratio of the mass functions to their mean, same color code. Bottom: Ratio of the standard deviation of the mass function to their mean, same color code.}
\label{fig:massfunc}
\end{figure}

Although the fixing process is not expected to impact the simulations at the dark matter halo level \citep{2018ApJ...867..137V}, list of halos are extracted from the different simulations for comparisons. Figure \ref{fig:massfunc} shows the mass functions within a 160~\hMpc\ radius sphere centered on the middle of the box for constrained paired (fixed) fields. As expected, there is no difference between the residual cosmic variances obtained with the constrained paired fixed and random fields. Note that again the mean mass function is not affected by the fixing process and that pairing is again important to obtain unbiased mean and residual cosmic variance.\\
 
 On the halo per halo basis, Virgo, Centaurus and Coma counterparts are identified in the different simulations as the unique halos, within a given region, massive enough to be considered as clusters. At this level, comparing the positions (x,y and z coordinates) and masses of the halos, the profits of constrained paired (fixed) fields over constrained non-paired fields are less obvious. The intrinsic scatter of the masses and positions of the total ensemble of Virgo, Centaurus and Coma halos from the constrained paired (fixed) fields is very similar to that of the halos in the constrained non-paired fields. Table \ref{Tbl:2} reports the mean mass values, their minimum, maximum and scatter as well as the mean X, Y and Z supergalactic positions in the six sets of constrained Virgo halos. The scatters are comparable, displaying no evidence of a clear decrease/increase in scatter when using paired (fixed) fields and no clear relations between the mean mass and position values. Results are similar for Centaurus and Coma. 
 
\begin{table*}
\begin{center} 
\begin{tabular}{ccccccccccc}
\hline
 \hline
&CRR$^+$  &&& CRR$^-$ & && CRR\\
               Virgo: $\langle$M$\rangle$&       5.62e+14  &  $\pm$(6.40e+13) & &       5.64e+14&  $\pm$(5.95e+13) & &      5.61e+14 &  $\pm$(6.88e+13)\\
                     M$_{min}$&       4.06e+14 &   & &       4.51e+14 &   & &       4.06e+14 &   & &\\
                     M$_{max}$&       7.09e+14 &   & &       7.07e+14 &   & &       7.09e+14 &   & &\\
      &&&&&&&\\               
   $\langle$SGX$\rangle$   &          -5.72 &     $\pm$(0.61) & &          -5.57 &      $\pm$(0.57) & &          -5.87 &      $\pm$(0.61)\\
   $\langle$SGY$\rangle$   &           6.37 &      $\pm$(0.60) & &           6.28 &      $\pm$(0.58) & &           6.45 &      $\pm$(0.61)\\
   $\langle$SGZ$\rangle$    &           3.52 &      $\pm$(0.74) & &           3.66 &      $\pm$(0.73) & &           3.38 &      $\pm$(0.73)\\
       &&&&&&&\\  
                     \hline
                           &&&&&&&\\    
&CFR$^+$  &&& CFR$^-$ & && CFR\\
               Virgo: $\langle$M$\rangle$&        5.57e+14  &  $\pm$(6.58e+13 ) & &       5.65e+14  &  $\pm$(4.66e+13) & &       5.50e+14 &  $\pm$(8.04e+13)\\
                    M$_{min}$ &       4.06e+14 &   & &       4.23e+14 &   & &       4.06e+14 &   & &\\
                     M$_{max}$&      7.70e+14 &   & &       6.38e+14 &   & &       7.70e+14 &   & &\\
        &&&&&&&\\    
    $\langle$SGX$\rangle$    &          -5.73 &      $\pm$(0.58) & &          -5.63 &      $\pm$(0.55) & &          -5.84 &      $\pm$(0.59)\\
    $\langle$SGY$\rangle$    &           6.38 &      $\pm$(0.58) & &           6.32 &      $\pm$(0.57) & &           6.43 &      $\pm$(0.59)\\
    $\langle$SGZ$\rangle$    &           3.52 &      $\pm$(0.69) & &           3.64 &      $\pm$(0.68) & &           3.40 &      $\pm$(0.69)\\
 \hline
\hline
\end{tabular}
\end{center}
\vspace{-0.25cm}
\caption{Mean masses in $\hmsun$, as well as their minimum, maximum and scatter, of the Virgo halos in different sets of constrained simulations. Mean X, Y, Z supergalactic coordinates and their standard deviations are given in \hMpc.}
\label{Tbl:2}
\end{table*}

\section{Conclusion}

To discriminate real tensions between observations and the standard cosmological model, revealed by recently reached precision cosmology, from a lack of accuracy, all possible kinds of systematics affecting our measurements must be considered. Among these potential actors of biases, our local environment produces effects of the order of the precision we expect to reach with future surveys, like those we will obtain for instance with the Euclid mission but also with the Large Survey Synoptic and 4-meter Multi-Object Spectroscopic Telescopes.  Mapping completely and precisely the local Universe is thus back on the front stage. With this renewed interest for `Near Field Cosmology' or the study of the local Universe as a whole, the region called local became as large as $\sim$300-400\hMpc. 

Cosmological simulations are now combined with detailed local observations in an attempt to achieve a fully complete picture of the local distribution of matter in order to understand it and its biasing effects. This effort gave rise to the development of initial conditions constrained by local observations. These initial conditions result in simulations that resemble the local Universe at redshift zero for a one to one comparison exercise almost free of cosmic variance.\\

However, these simulations present a common pitfall that is they represent plausible models of the local Universe but not the local Universe model. The residual cosmic variance between the different realizations of the local Universe implies the need for hundreds of runs before drawing sensible conclusions and their associated uncertainties.\\

This paper diverts the `fixed-paired' technique from its original use in an attempt to propose an alternative to the ultimate solution that would imply overcoming both the non-linearities of the problem and the noisy observational data available only today in a limited volume as well as the limited size and resolution of the box to get \emph{the} local Universe simulation. The constraining algorithm applied to paired fixed fields rather than random fields permits obtaining simulations to efficiently \#1 disentangle the different responsibilities leading to this residual cosmic variance, namely evaluate the large scale fraction that is completely unaffected by the constraints, \#2 estimate the uncertainty or residual cosmic variance on a local Universe simulation and \#3 provide a mean estimate of an ensemble of local Universe simulations:
\begin{itemize}
\item By construction, constrained simulations all resemble the local Universe. They differ solely by the random realization to which the constraints are combined to build initial conditions. Their cosmic variance is thus reduced by a factor 2 to 3 in the inner part of the box where most of the constraints are with respect to random simulations \citep{2018MNRAS.478.5199S}. Estimating the residual cosmic variance requires hundreds of these constrained simulations.
\item Constrained fixed simulations differ from the typical constrained simulations by being built from fixed realizations rather than random realization, namely the amplitudes of the modes are fixed. Their intrinsic scatter is found to be smaller in terms of the power spectrum but only by up to 20\% for the large scale modes. These simulations show that most of the residual cosmic variance is due to correlations between large scales and local observational constraints thus that these scales are constrained up to 80\% by the local data. Only 20\% is purely random.
\item Constrained paired simulations, that differ from one another only by the realization, where one is the exact opposite of the other, are excellent proxy of the mean of several constrained simulations. They give access to an optimal measurement of the residual cosmic variance in the sense that it is not biased like with a random subset of constrained simulations. The uncertainty on simulated structures is directly given by the variance between constrained simulations obtained with two paired fields of the same pair without the requirement for hundreds of runs.
\item Constrained paired fixed simulations gather both profits without adding a systematic. The fixing process biases neither the mean density fields nor properties of velocity fields (monopole and dipole).
\end{itemize}

The utility of the constrained paired fixed simulations relies on the growing interest in the geometric mean of constrained fields and on evaluating the large scale validity of local Universe simulations. Both points require hundreds of constrained simulation runs to derive unbiased variance and mean values and thus draw sensible conclusions. The geometric mean of constrained fields has already been used in \citet{2018NatAs...2..680H} to determine the luminosity-bias in the quasi-linear regime. In this paper, the geometric mean is that of a small number of constrained fields without using paired fields. We thus claim that the geometric mean of two constrained paired fields of the same pair is at least equally closer, if not more, to the true geometric mean than that of a small number of constrained fields obtained with completely independent seeds (random realizations). It is thus completely appropriate to determine the quasi-linear local density field at a considerably smaller computational cost. Another example is that of  \citet{2017MNRAS.471.3087S} who used a large number of constrained realizations to derive the probabilities of structures in the zone of avoidance. The two constrained paired fixed fields of the same pair provide now a faster and efficient way of obtaining these probabilities at a much smaller computational cost. Additionally, their standard deviation provides the residual cosmic variance or uncertainty on any local Universe simulation disentangling the uncertainty part due to completely unconstrained large scales to those partly constrained.\\

More broadly, the constrained paired fixed simulations will be extremely useful in determining the accuracy of the simulations in reproducing the local Large Scale Structure, source of foreground effects on background large scale surveys and on the cosmic microwave background. However, when it comes to precisely study the local cluster-size halos, statistical studies are still of use to determine their average properties.  Still, for the most constrained halos, the scatter is already small and studying at high resolutions with the zoom-in technique a constrained halo in one of the constrained simulation should already be a good proxy for the observed cluster.

\section*{Acknowledgements}
The author would like to warmly thank the referee for their useful comments that helped clarify the paper. JS would like to thank A. Pontzen, H. Peiris, M. Rey and J. Blaizot for useful conversations. This work was supported by the `Programme National Cosmologie et Galaxies' (PNCG) of CNRS/INSU with INP and IN2P3, co-funded by CEA and CNES. The author gratefully acknowledges the Gauss Centre for Supercomputing e.V. (www.gauss-centre.eu) for providing computing time on the GCS Supercomputers SuperMUC at LRZ Munich.

\section*{Appendix}
Additional initial conditions as described in Table \ref{Tbl:1} are of interest for other studies.  They were prepared and run from redshift 60 to redshift 0 using gadget \citep{Springel2005} with 256$^3$ dark matter particles (particle mass 6.4$\times$10$^{11}\hmsun$) in 500~\hMpc\ boxes within the Planck cosmology framework \citep[$\Omega_m$=0.307, $\Omega_\Lambda$=0.693, H$_0$=67.77~km~s$^{-1}$~Mpc$^{-1}$, $\sigma_8$~=~0.829,][]{2014A&A...571A..16P}. \\

\begin{figure*}
\vspace{-1.8cm}
\includegraphics[width=0.85 \textwidth]{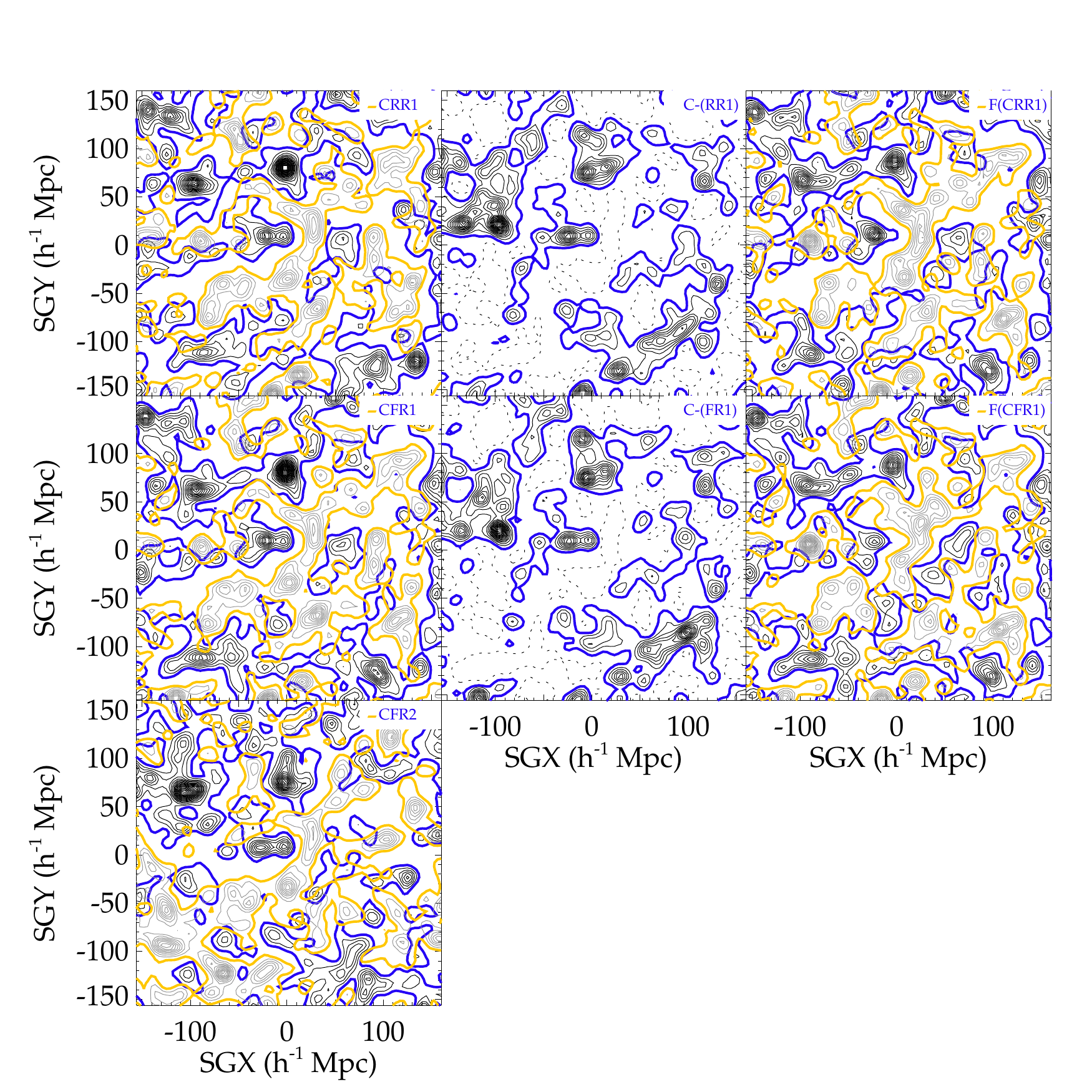}
\caption{XY supergalactic slices of the density fields of different simulations. See Table \ref{Tbl:1} for an explanation of the abbreviations. Solid black contours show overdensities. Dashed contours stand for the underdensities. Blue and yellow colors represent the mean field of two paired fields respectively.}
\label{fig:add1}
\end{figure*}

\begin{figure*}
\vspace{-0.5cm}
\includegraphics[width=0.5 \textwidth]{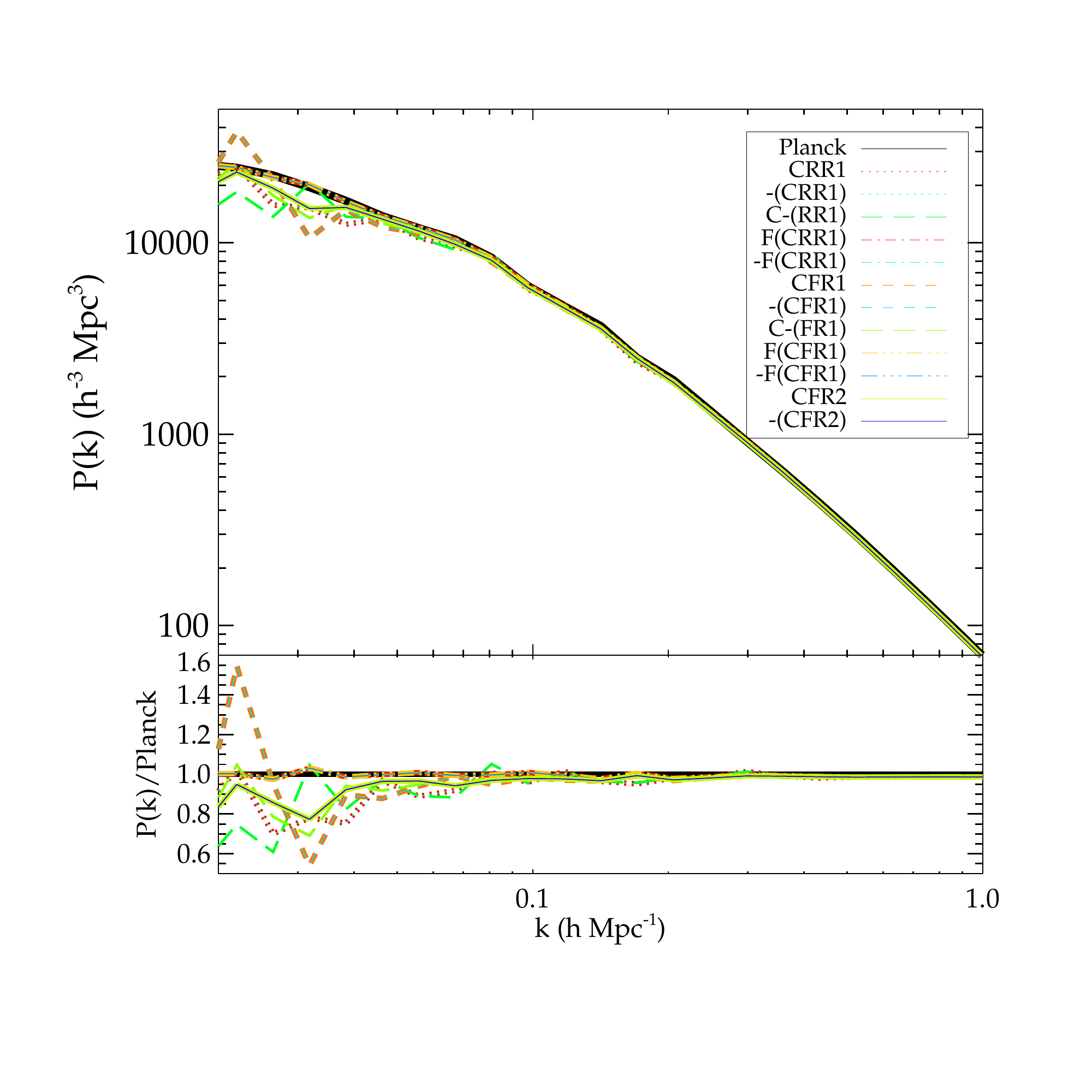}\hspace{-1cm}
\includegraphics[width=0.5 \textwidth]{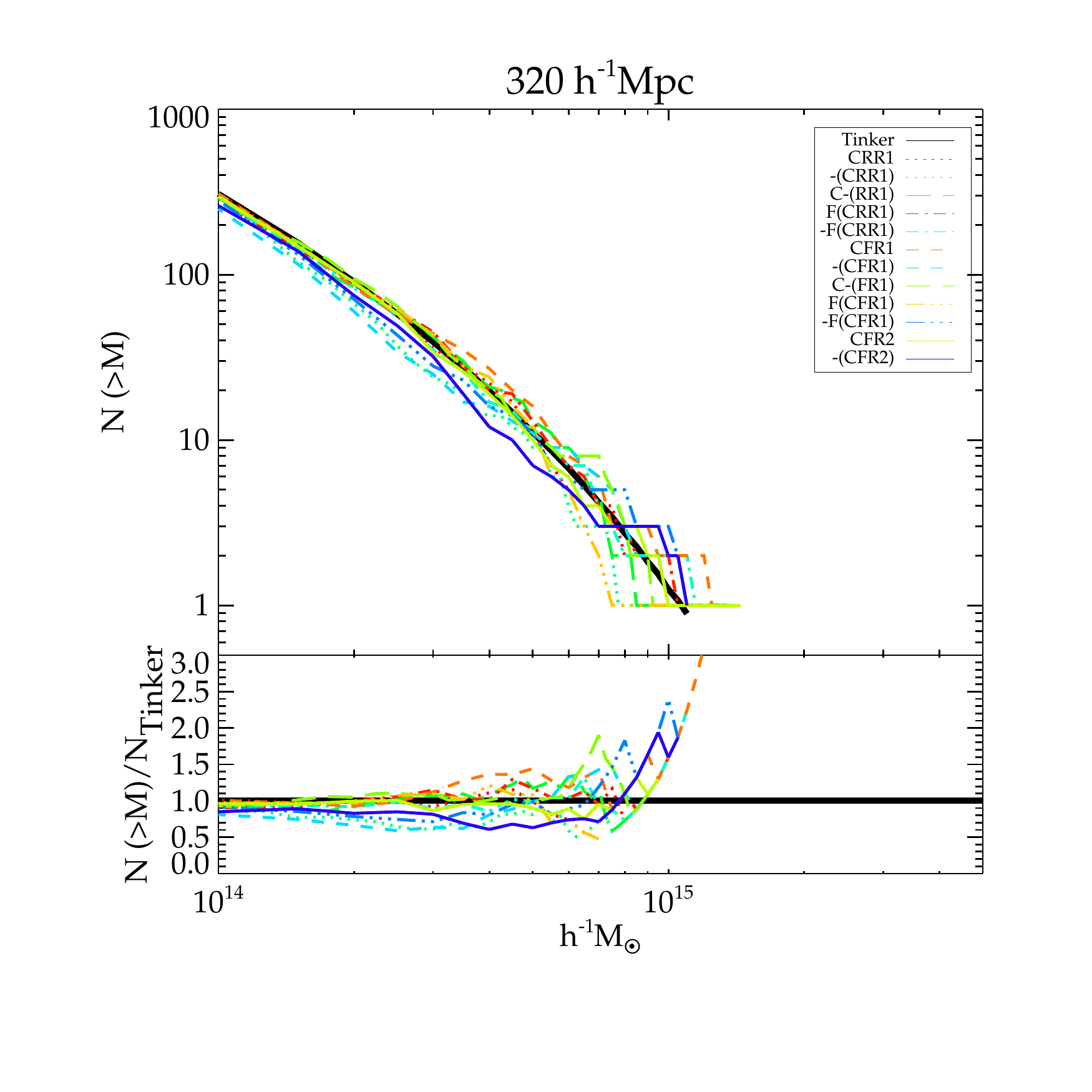}
\vspace{-1cm}
\caption{Top: power spectra (left) and mass functions (right) of different constrained simulations. See Table \ref{Tbl:1} for an explanation of the abbreviations. Bottom: power spectra (left) and mass functions (right) divided by Planck power spectrum (right) or Tinker mass function (left), same color code.}
\label{fig:add2}
\end{figure*}

Figure \ref{fig:add1} shows the XY supergalactic slices of the density fields of different constrained simulations with short names given at the top right corner of each small panel. Their power spectrum and mass function are visible in Figure \ref{fig:add2}. Explanations for the short names are given in Table \ref{Tbl:1}.  \\

Interestingly, fixing the fields of the initial conditions after constraining (F(CRR) or F(CFR)) reveals that the pair Virgo-Centaurus clusters disappear from the resulting simulations, leaving room for only one density peak. This suggests that the pair Virgo-Centaurus is not representative of a common environment. It indicates that our close environment is not an average environment but really suffers from the cosmic variance. Additionally, it is also clearly visible that while Virgo-Centaurus are really well constrained, Coma, Shapley and Perseus are less constrained since the resulting density field varies between simulations based on two paired fields of the same pair. Note that there is no obvious visual difference between fields smoothed at 5~\hMpc\ obtained with a random realization and its fixed counterpart. Finally, density peaks in paired constrained simulations, i.e. initial conditions are paired after constraining, could help us understand and study local voids as their counterparts.

\label{lastpage}
\bibliographystyle{mnras}

\bibliography{biblicomplete}
 \label{lastpage}
\end{document}